\documentstyle[12pt,epsf,worldsci]{article}
\begin{document}
\title{QUARK DYNAMICS and FLUX TUBE STRUCTURE \\
 in $q\bar{q}$ BOUND STATES }
\author{Nora Brambilla\\
{\em Institut f\"ur Theoretische Physik, Boltzmanngasse 5, A-1090,
 Vienna, Austria}}
\maketitle
\setlength{\baselineskip}{2.6ex}

\vspace{0.7cm}
\begin{abstract}
I discuss the bound state quark dynamics focusing on the nonperturbative interaction and
  the flux tube structure.
\end{abstract}
\vspace{0.7cm}

\section{Introduction}

\begin{figure}[htb]
\begin{centering}
\vskip -0.1truecm
\makebox[0truecm]{\phantom b}
\epsfxsize=7.5truecm
\epsffile{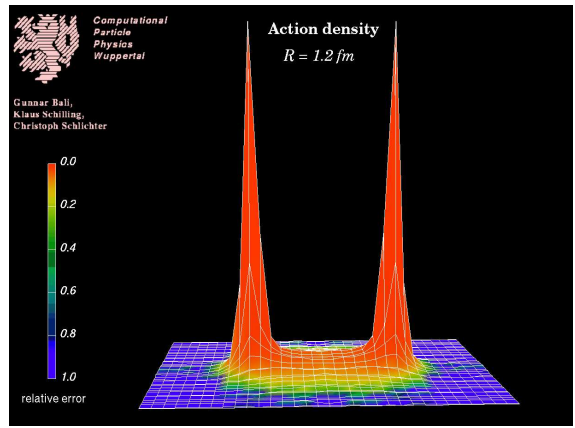}
\vskip -0.1truecm
\caption[x]{Lattice evidence of the interquark flux tube.
Figure provided by G. Bali, see\cite{baliflux,bali}.}
\label{figuno}
\end{centering}
\end{figure}
\vspace{0.3cm}
Confinement and chiral symmetry breaking are the two main facts in  low-energy QCD and show up 
explicitly in the spectrum of quark bound states.
Confinement implies that the nature of the  quark dynamics has to change strongly   
 in dependence on the scale of energy or distance associated with the interaction.
Let us consider  the interaction between a quark and an antiquark. At short
 quark-antiquark 
distance $r$, the interaction is dominated by the  one gluon exchange (dipole configuration)
 and is obtained by a perturbative expansion in $\alpha_{\rm s}$.
 With increasing  distance,   the interaction becomes linear
 in $r$ (flux tube configuration). The formation of a flux tube of constant energy density $\sigma
\simeq  1{\rm GeV}/ {\rm fm}$ 
(the so called string tension) confining the quark and the antiquark
 to remain bound, is the 
most striking feature  
of confinement and is inherently nonperturbative.
We have hints about this flux tube   from the phenomenological spectrum and 
the Regge  trajectories. Phenomenological   potential models include  interactions rising with the
 quark-antiquark  distance. 
Dual models explain this feature
as a chromoelectric flux tube formation due to a dual Meissner effect, 
 the QCD vacuum behaving like a t'Hooft-Mandelstam superconductor. Confinement is 
 then due to monopoles condensation. \par
Lattice studies  of the electric and magnetic fields strength distributions 
 around a static quark-antiquark pair yield direct evidence  and a 
detailed picture  of the string formation 
(see Fig.\ref{figuno}). 
Lattice data provide us with: a quark-antiquark static potential linear 
for $r> 0.2\,  {\rm fm}$; an interquark flux tube structure with a transversal rms width ranging 
between  
 $ 0.5 \, {\rm fm} $ and $0.75\, {\rm fm}$;
   positive tests of monopole condensations; Abelian projected electric fields 
and currents distributions in agreement with 
 the dual Ginzburg Landau predictions\cite{bali,pol}.\par 
The lattice remains the most suitable tool to obtain nonperturbative information.
However,  most of these appear as independent and uncorrelated  measurements.
On the other side we have many phenomenological models  implementing confinement.
Finally, high order perturbative calculations  are also  available.
In order to gain some insight in the mechanism of confinement and  flux tube formation
 we need a model-independent, exact (in the sense that it contains
perturbative and nonperturbative physics and is valid at any scale), gauge-invariant
and {\it systematic} approach to the quark interaction. It helps to have a 
 formulation which is  closely related to the lattice,  allows analytic calculations and
is physically transparent.\par
In Secs. 2 and 3 
we show that this program is realizable in the heavy quark case since 
a small expansion parameter (the quark velocity $v$) still exists. 
The result obtained is model independent and suitable for lattice evaluation. 
An analytic evaluation is  possible  once a QCD vacuum model is considered.
Therefore it is possible to test models of confinement  in an unambiguous and 
direct relation with the phenomenological and lattice data.\par
We know that in the low energy region new degrees of freedom become relevant.
The flux tube formation hints to the fact
that dynamical gluon effects are vital when attempting to incorporate
nonperturbative interaction, indeed the gluonic degrees of freedom are condensed into 
stringlike flux tube. Hence, the  flux tube   should determine the nonperturbative
bound state  quark dynamics. On one side, this  means that also excited states of the interquark
 string has to be considered. This leads to the consideration 
of  hybrids states  and was discussed by Michael\cite{michael}
at this Conference. On the other side, this means  that even when considering the ground state of the
 interquark 
glue, the flux tube energy should be taken into account properly. 
Moreover, to understand the transition region, we should be able to control precisely how
the collective behavior of gluons modifies and alters single gluon exchange.
\par
In Secs. 4, 5 and 6 we discuss the heavy quark dynamics in 
 QCD vacuum models that reproduce the flux tube structure,
in  comparison with the lattice data and 
with the ``$ {1/ Q^4}$'' models extensively used 
in the literature. We find that all the   'flux tube models' 
 predict the same form
for the heavy quark interaction also  in the intermediate region.   
We show that the Gaussian dominance of the Wilson loop average has to be 
related   with the Abelian dominance of infrared QCD establishing a connection with 
the dual superconductor mechanism.  \par
In Secs. 7 and 8 we extend our discussion to bound systems involving at least one light quark.
Here the spontaneous  chiral symmetry symmetry breaking emerges in an interplay with confinement.

\section{Gauge-Invariant approach to quark bound states}

  We consider a  gauge invariant quark-antiquark singlet state\footnote{Actually
 this state is not 
an eigenstate of the Hamiltonian. It serves as a trial state
 to extract the energy of the lowest eigenstate 
having a non-vanishing projection on $\vert \phi\rangle$. This is equivalent to say that the 
contribution of the string is negligible in the limit $T\to \infty$, which will become more 
apparent when dealing with the Wilson loop.} 
\begin{equation}
\vert \phi_{\alpha \beta}^{lj} \rangle \equiv {\delta_{lj}\over \sqrt{3}}
\bar{\psi}^i_\alpha(x) U^{ik}(x,y,C) \psi^k_\beta(y) \vert 0\rangle
\label{statgaug}
\end{equation}
where $i,j,k,l=1,\dots 3$ are color indices  and the Schwinger string line $U$  has been inserted 
to ensure gauge-invariance 
\begin{equation}
U(x,y;C) =P \exp \{ i g \int_{y,C}^x A_\mu(z) dz^\mu \} .
\label{string}
\end{equation}
At a time $t=0$, a quark and an antiquark are created, 
  interact while propagating for a time time $t=T$ at which they are annihilated.
Then, the quark-antiquark interaction is contained in 
the gauge-invariant four-point Green function
($x_j=({\bf x}_j,T)$, $y_j=({\bf y}_j,0)$)
\begin{equation}
\!\!\!\!\!\!\! G(x_1,x_2,y_1,y_2)  =  {1\over Z} \int {\cal D}A \,
 Tr  S(x_2,y_2; A)  U(y_2,y_1)  
  S(y_1,x_1;A)  U(x_1,x_2)   e^{i S_{\rm YM}}
\label{green}
\end{equation}
where we neglected  the  fermionic determinant and the annihilation terms\footnote{Since
 we neglect the fermionic determinant
 we work in the quenched approximation. In the case of heavy quarks however we can 
 introduce into the effective action the expansion of the determinant\cite{vairo}.},
$Tr$ is the color trace and  $S(y,x;A)$ is the quark propagator in the external field $A$.
To find a  closed expression for the Green function in Eq.(\ref{green}) is quite a 
formidable task. However, the situation becomes simpler  if we consider the case of heavy quarks. 
In the static limit $m\to \infty$  the equation for the static quark propagator 
\begin{equation}
 (i\gamma_0 D_0 -m)S_0(x,y;A)= \delta^4(x-y)
\label{dirstat}
\end{equation}
is solvable in closed form \cite{brown,eichten} and we find \cite{wilson,brown,peskin}
\begin{equation}
G(T)   {\buildrel {m_j\to\infty }\over \longrightarrow } 
\delta^3({\bf x}_1-{\bf y}_1) \delta^3({\bf x}_2-{\bf y}_2)
 e^{-i(m_1+m_2) T}  \langle W(\Gamma_0)\rangle
\label{limit}
\end{equation}
where
\begin{equation}
\langle W(\Gamma_0)\rangle  = {1\over Z} \int {\cal D}A \,
 Tr P  \exp \big \{ {ig \oint_{\Gamma_0} dz^\mu A_\mu (z)} \big \}
 \exp \{ ig S_{\rm YM}\big \}
\label{wilstat}
\end{equation}
is the static Wilson loop average, the loop $\Gamma_0$ being a rectangle.
 From the Feynman--Kac formula and Eq. (\ref{limit}) we get 
\begin{equation}
V_0(r)  = \lim_{T\to \infty} {i\over T } \ln   \langle W(\Gamma_0) \rangle .
\label{stat}
\end{equation}
The static quark-antiquark potential, {\it if exists},
 is given in terms of the static Wilson loop average. Notice that this is an exact 
expression that contains both the perturbative and the nonperturbative dynamics. Indeed, 
 no expansion  in the coupling constant has been performed.
 Here and in the following sections
we implicitly assume the existence of the potential and we postpone this issue  
to Sec 6.
The spin-dependent ${1/ m^2}$ corrections to the potential were calculated in 
\cite{eichten,gromes} considering the first correction to Eq. (\ref{dirstat}) and 
evaluating it on the zero order exact solution. The result is 
  given in terms of 
 vacuum expectation value of (color) electric and magnetic field  insertions in the 
static Wilson loop and controls the fine and hyperfine separation in quarkonia. \par
However there were   some residual difficulties:
\begin{itemize}
\item{} The kinetic energy terms was not considered\footnote{The static quark propagator is
the zero order solution. This is similar to the HQET approach.}
lacking in consistency with the virial  theorem.
\item{} The spin-dependent potentials appeared as an expansion in ${1/ m^2}$  completely 
missing  the terms in $\ln m$ obtained  at one loop 
in the $S$ matrix perturbative calculation of the interaction\cite{gupta}.
\item{} The calculation of the nonperturbative contributions was based  on several assumptions 
and there was  no clear and unambiguous procedure to calculate the nonperturbative 
behavior of the v.e.v. of the field strength insertions in the static Wilson loop.
\item{} It was  not clear when a potential description was  holding.
\end{itemize}
\par
The solution is {\it to  disentangle the different scales of the bound system} ($m$, $p=m v$,
 $E=m v^2$, $v$ being the heavy quark velocity) and to establish
 a {\it systematic }, {\it unambiguous} and {\it exact} expansion procedure. This will be 
reported in the next section.

\section{Analytic closed expression for the heavy quark interaction}

In the  heavy quark systems, the existence of an 
expansion parameter (the inverse of the mass $m$ in the Lagrangian and the 
velocity $v$ of the quark as a dynamical defined power counting parameter) 
makes possible to establish a systematic expansion procedure.
 The tool is provided by NRQCD \cite{nrqcd}. 
This is an effective theory equivalent to QCD and obtained from 
QCD by integrating out the hard energy scale $m$. The Lagrangian comes from the 
original QCD Lagrangian via a Foldy--Wouthuysen transformation. The 
ultraviolet regime of QCD (at energy scale $m$) is perturbatively 
encoded order by order in the coupling constant $\alpha_{\rm s}$ 
in the matching coefficients which appear in front of the new operators 
of the effective theory. This ensures the equivalence between the effective  
theory and the original one at a given order in $1/m$ and $\alpha_{\rm s}$. 
At order $1/m^2$ the NRQCD Lagrangian describing a bound state between 
a quark of mass $m_1$ and an antiquark of mass $m_2$ is \cite{manohar,pineda}
\begin{eqnarray}
& &  L = \psi_1^\dagger\!\left(\!iD_0 + c^{(1)}_2 {{\bf D}^2\over 2 m_1} + 
c^{(1)}_4 {{\bf D}^4\over 8 m_1^3} + c^{(1)}_F g { {\bf \sigma}\cdot {\bf B} \over 2 m_1} \right.
 c^{(1)}_D g { {\bf D}\!\cdot\!{\bf E} - {\bf E}\!\cdot\!{\bf D} \over 8 m_1^2} 
  \nonumber \\
& &+ \left.  i c^{(1)}_S g {{\bf \sigma}
 \!\!\cdot \!\!({\bf D}\!\times\!{\bf E} - {\bf E}\!\times\!{\bf D})
\over 8 m_1^2} \!\right)\psi_1
+ \hbox{\, antiquark terms}\, (1 \leftrightarrow 2)
+ {d_1\over m_1 m_2} \psi_1^\dagger \psi_2 \psi_2^\dagger \psi_1 
\nonumber \\
& &  + {d_2\over m_1 m_2} \psi_1^\dagger {\bf \sigma} \psi_2 \psi_2^\dagger {\bf \sigma} \psi_1
+ {d_3\over m_1 m_2} \psi_1^\dagger T^a \psi_2 Q_2^\dagger T^a \psi_1 
+ {d_4\over m_1 m_2} \psi_1^\dagger T^a {\bf \sigma} \psi_2 
\psi_2^\dagger T^a {\bf \sigma} \psi_1.
\label{nrqcd}
\end{eqnarray}
This is the relevant Lagrangian in order to calculate the bound state 
energies up to order $O(v^4)$.  The coefficients $c^{(j)}_2$, $c^{(j)}_4$, ... 
are evaluated at a matching scale $\mu$ for a particle of mass $m_j$. 
 The  matching coefficients are $1$ or $0$ at the tree level.
At one loop they contain terms in  $\ln m_j$ which establish
 \cite{chen} the agreement of the present
calculation with the result for the perturbative  interaction obtained at one loop 
 in  the $S $ matrix formalism\cite{gupta}. For simplicity  in the following we consider the tree level 
values \footnote{For a discussion of the matching coefficients, the 4-fermions interaction
 and   the power in $v$ of the  operators 
 in (\ref{nrqcd}) see \cite{vairo}.}.\par
\begin{figure}[htb]
\begin{centering}
\vskip -0.3truecm
\makebox[2truecm]{\phantom b}
\epsfxsize=8.5truecm
\epsffile{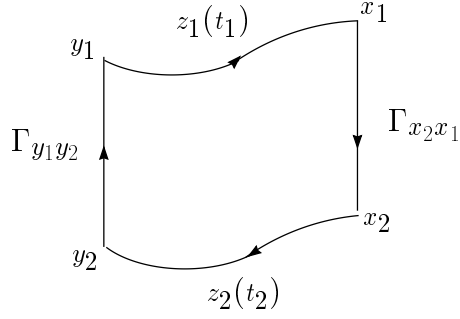}
\vskip -6.0truecm
\caption[x]{Distorted Wilson loop.}
\label{figdue}
\end{centering}
\end{figure}

With Eq. (\ref{nrqcd}) we have reduced the heavy quark dynamics to a nonrelativistic Schr\"odinger 
theory for the heavy quark (antiquark) with Pauli propagator $K$ coupled to the usual 
relativistic field theory for light quark (omitted here, quenched appr.) and gluons.
This result is especially useful for calculation on coarse lattice \cite{nrqcd}.
However, here we are looking for analytic results. The strategy is the following \cite{vel,hugs}.
At the $O(v^2)$  the Pauli propagator satisfies the equation
\begin{eqnarray}
& & i \frac{\partial}{\partial x^0} K_j =  
 \left[ m_j +\frac{1}{2m_j} ({\bf p}_j - g{\bf A})^2 \right. 
  -  \frac{1}{8m_j^3} ({\bf p}_j - g{\bf A})^4 + gA^0
\label{schro} \\
& &\ - \frac{g}{m_j}{\bf S}_j \cdot {\bf B}   
  - \frac{g}{8m_j^2} (\partial_i E^i - ig [A^i,E^i]) 
 \left. +  \frac{g}{4m_j^2} \varepsilon^{ihk} S_j^k
\{(p_j -gA)^i,E^h\} \right]  K_j 
\nonumber
\end{eqnarray}
and therefore admits a path integral representation
\begin{equation}
K_j(x,y|A)= \int {\cal D} [{\bf z}_j, {\bf p}_j] \, {\rm T}_s \, 
e^{  i \int 
dt \, [{\bf p}_j \cdot \dot{{\bf z}}_j -  H ]  } 
\label{path}
\end{equation}
where $H$ is the Hamiltonian appearing in (\ref{schro}).
Starting, as in Sec. 2, from the $q\bar{q}$ Green function (\ref{green}), we can substitute the quark
Dirac propagator with the Pauli propagator and use Eq. (\ref{path}) 
   to obtain the  closed path integral
representation
\begin{eqnarray}
& & \!\!\!\!\!\!\!\!\!\!\!\!\!\!\!\  G(T)  \longrightarrow K(T) =
\int {\cal D}  [{\bf z}_1, {\bf p}_1]\int {\cal D} [{\bf z}_2, {\bf p}_2] 
 \exp\left\{i\!\int\! dt\, \sum \left[{\bf p}_j \cdot \dot{{\bf z}}_j -m_j-
\frac{{\bf p}^2_j}{2m_j}+\frac{{\bf p}^4_j}{8m_j^3}\right] \right\} \nonumber \\
& & 
\times \, \left\langle \frac{1}{3}{\rm Tr \, T_s \, P} \exp\left\{ig\oint_{\Gamma}  dz^{\mu} \,
A_{\mu}(z) \right.\right. +
\label{rapr}\\
& & \sum_{j=1}^2\frac{ig}{m_j}      \int_{\Gamma_j} dz^{\mu} 
\big (S_j^l \hat{F}_{l{\mu}}(z)    - 
\frac{1}{2m_j}
S_j^l\varepsilon^{lkr}p_j^k F_{{\mu}r}(z) 
  \left.\left. - \frac{1}{8m_j} D^{\nu}F_{{\nu}{\mu}}(z) \big ) \right\} 
\right\rangle 
\nonumber
\end{eqnarray}  
with $ \langle f[A]\rangle\equiv
{1\over Z} \int {\cal D} A \,f[A] \exp\{ i S_{\rm YM}\}$ , $T_{\rm s}$ spin ordering and 
$P$ path ordering. Notice that now the kinetic energy of the quark is explicitly considered 
and the contour integral in $A_\mu$ and $F_{\mu \nu}$ is extended to the distorted 
Wilson loop in Fig. \ref{figdue}. Indeed, due to the presence of the path integral sum, we are considering 
any possible trajectory for the quark (antiquark) at variance with the static path of the
previous section.\par
\vspace{0.3cm}
Then, the potential exists if we can make the following identification in (\ref{rapr})
\begin{equation}
 \left\langle \frac{1}{3}{\rm Tr} \cdots\right \rangle =
{\rm T_s} \exp\left[ -i\int  dt \,
 V({\bf z},{\bf p},{\bf S}) \right] .
\label{exist}
\end{equation}
From Eqs.(\ref{rapr}) and (\ref{exist}) we find \cite{vel}
\begin{eqnarray}
&& 
\int_{t_{\rm i}}^{t_{\rm f}} dt V_{{\rm Q} \bar{{\rm Q}}} =
 i \log \langle W(\Gamma) \rangle 
\label{potfin}\\
& &   - \sum_{j=1}^2 \frac{g}{m_j} \int_{{\Gamma}_j}dz^{\mu} 
\left. \Big ( S_j^l \, 
\langle\!\langle \hat{F}_{l\mu}(z) \rangle\!\rangle\right.  
  -\frac{1}{2m_j} S_j^l \varepsilon^{lkr} p_j^k \, 
\langle\!\langle F_{\mu r}(z) \rangle\!\rangle    - \left. \frac{1}{8m_j} \, 
\langle\!\langle D^{\nu} F_{\nu\mu}(z) \rangle\!\rangle \right. \Big ) 
\nonumber \\
& &  -  \sum_{j,j^{\prime}=1}^2 \frac{ig^2}{2 m_jm_{j^{\prime}}}
{\rm T_s} \int_{{\Gamma}_j} dz^{\mu} \, \int_{{\Gamma}_{j^{\prime}}} 
dz^{\prime\sigma} \, S_j^l \, S_{j^{\prime}}^k
  \left( \, 
 \langle\!\langle \hat{F}_{l \mu}(z) \hat{F}_{k \sigma}(z^{\prime})
\rangle\!\rangle  -\right. 
 \left.\langle\!\langle \hat{F}_{l \mu}(x) \rangle\!\rangle
\, \langle\!\langle \hat{F}_{k \sigma}(x^{\prime}) \rangle\!\rangle  \right) 
\nonumber
\end{eqnarray}
where $
 \langle\!\langle f(A) \rangle\!\rangle \equiv \langle f(A) W(\Gamma)\rangle/ \langle W(\Gamma)\rangle$
  is the  vacuum expectation value in presence of a quark--antiquark pair (i.e. the Wilson loop).
Eq. (\ref{potfin}) is   the quark-antiquark
potential at order $v^2$ of the systematic expansion in $v$. No expansion in the coupling 
constant has been performed.
 The result is physically 
transparent. In Eq. (\ref{potfin}) 
 the first term contains the static and the velocity-dependent 
potential, the second  is the magnetic interaction, the third is the  Thomas precession, the 
fourth is  the Darwin term and the last one is the spin-spin interaction.
All the dynamics is contained in the Wilson loop and in the v.e.v. of field strength in 
presence of the Wilson loop, hence is pure gluodynamics.
Had we worked in QED, we would obtain the same result as in (\ref{potfin}). Notice that 
the coupling underlying (\ref{potfin}) is the vectorial coupling of the Dirac equation.
The difference between QED and QCD  is contained in the
behavior of the  v.e.v. of field strengths 
and Wilson loop. The non linear dynamics of QCD determines the nonperturbative behavior 
of these v.e.v and accounts for the difference with QED. \par
Varying the quark path 
 $z_j^\mu (t) \rightarrow  z_j^\mu (t) + \delta  z_j^\mu (t)$, the Mandelstam formula can be 
 obtained 
\begin{eqnarray}
& &\ g\langle\!\langle F_{\mu\nu}(z_j) \rangle\!\rangle = (-1)^{j+1}
{\delta i \log \langle W(\Gamma) \rangle \over \delta S^{\mu\nu} (z_j)} \nonumber  \\
& &\ g^2 \!\left(\!\langle\!\langle F_{\mu\nu}(z_1) F_{\lambda\rho}(z_2) \rangle\!\rangle 
- \langle\!\langle F_{\mu\nu}(z_1) \rangle\!\rangle 
  \langle\!\langle F_{\lambda\rho}(z_2) \rangle\!\rangle \!\right) 
  = - i g {\delta\over \delta S^{\lambda\rho}(z_2)} 
\langle\!\langle F_{\mu\nu}(z_1) \rangle\!\rangle
\nonumber
\end{eqnarray}
with $ \delta S^{\mu\nu} (z_j) =  dz_j^\mu \delta z_j^\nu - dz_j^\nu \delta z_j^\mu $.
We conclude that 
to obtain the complete  quark-antiquark order $O(v^2)$ interaction (quenched)  no other 
assumptions are needed than the behavior of $\langle W(\Gamma) \rangle$:
{ \it given  $\langle W(\Gamma) \rangle$
 everything is analytically calculable.}
 On the other side, expanding the  average on the distorted Wilson loop $\Gamma$
in terms of the static Wilson loop $\Gamma_0$ we get expressions for the potentials
suitable for lattice evaluation \cite{potlat}.
Then, {\it the way to validation of  analytic models
 of the  QCD vacuum  via lattice  data and  phenomenological data is open}. 
The aim is to obtain  information as much as possible model-independent  on the nonperturbative dynamics.

\subsection{General form of the potential}

At the order $O(v^2)$ and with tree level matching coefficients
 the quark-antiquark 
interaction reads 
\begin{equation}
 V =  V_0  +  V_{\rm SD}+  V_{\rm VD} 
\label{defpot}
\end{equation}
with the spin-dependent interaction
\begin{eqnarray}
& & \!\!\!\!\!\!\!\!\!  V_{\rm SD} =
 {1\over 8} \left( {1\over m_1^2} + {1\over m_2^2} \right) +
 \Delta  \left[   V_0(r)  + V_{\rm a}(r)  \right] 
 \left( {{\bf L}_1 \cdot {\bf S}_1 \over 2 m_1^2} 
- {{\bf L}_2 \cdot {\bf S}_2\over 2 m_2^2}  \right) 
{1\over r} \left[  V^\prime_0(r)+ 2  V^\prime_1(r)  \right] \label{vsd}\\
& &  \!\!\!\!\!\!\!\!  +
{1\over m_1 m_2} \left( {\bf L}_1 \cdot {\bf S}_2 - {\bf L}_2 \cdot {\bf S}_1 \right) 
 {V^\prime_2(r)\over r} +
{1\over m_1 m_2}  \left( { {\bf S}_1\cdot{\bf r} \> {\bf S}_2\cdot{\bf r}\over r^2} 
- { {\bf S}_1 \cdot {\bf S}_2 \over 3}\right)  V_3(r)
+   {{\bf S}_1 \cdot {\bf S}_2\over 3 m_1 m_2}   V_4(r) 
\nonumber
\end{eqnarray}
where $
{\bf L}_j = {\bf r} \times {\bf p}_j$,
 and the velocity-dependent interaction is given by
($\Big\{~~\Big\}  =$ Weyl  ordering),
\begin{eqnarray}
& &  V_{\rm VD}=   {1\over m_1 m_2} 
\left\{ {\bf p}_1\cdot{\bf p}_2  V_{\rm b}(r) \right\} 
+    {1\over m_1 m_2} \left\{ \left( {{\bf p}_1\cdot{\bf p}_2\over 3} - 
{{\bf p}_1\cdot {\bf r} ~ {\bf p}_2 \cdot {\bf r} \over r^2}\right) 
 V_{\rm c}(r) \right\} \nonumber  \\
& & +  \sum {1\over m_j^2}
\left\{  p^2_j  V_{\rm d}(r) \right\}
+  \sum {1\over m_j^2}\left\{\left( {p^2_j\over 3} - 
{{\bf p}_j\cdot {\bf r} ~ {\bf p}_j \cdot {\bf r} \over r^2}\right) 
 V_{\rm e}(r) \right\} .
\label{vd}
\end{eqnarray}
The dynamics is contained in the $V_i(r)$ functions. These can be evaluated analytically 
via functional derivatives or expansion of the Wilson loop (once its nonperturbative 
behavior is assigned in a vacuum model), see \cite{mod,vel}
 or can be evaluated numerically on the lattice 
from v.e.v. of field strength insertions in the static Wilson loop, see\cite{potlat,balipot}.  
The spin-dependent potentials agree with the Eichten-Feinberg result \cite{eichten}. The velocity 
dependent potentials were obtained in \cite{vel,potlat} and {\it have to be taken into account} 
to be consistent at the order $v^2$ of the systematic expansion.
The lattice results are presented in Sec. 5  while in the next Section we summarize  the analytic
results obtained in various models of the QCD vacuum. The common thread of these models
is  reproducing the interquark flux tube structure. \par

\section{Infrared dynamics and flux tube structure}
The expression   (\ref{potfin}) for the heavy quark interaction is exact at the order of the expansion in 
$v$ considered. It is gauge-invariant and this allows us to insert approximations, i.e.
 vacuum models to get the nonperturbative behavior of the Wilson loop average. 
We retain the relevant configuration in the nonperturbative regime, which turns out to be 
a flux-tube like configuration.

\subsection{A simple model calculation}

The simplest model consist in assuming that the short and the long range contributions to the Wilson loop 
factorize and evaluating the last one 
 using an area law 
 (as suggested by the strong coupling expansion)
\begin{equation}
 \log \langle W (\Gamma) \rangle =  \log \big [ \langle W (\Gamma) 
\rangle^{\rm SR}\cdot 
 \langle W (\Gamma) 
\rangle^{\rm LR}\big ]
  = - \frac{4}{3} g^2 \oint_{\Gamma}dx^{\mu} \oint_{\Gamma} dy^{\nu} D_{\mu \nu} (x-y)  
 + \sigma S_{\rm min} + {C\over 2} {\cal P}.
\label{mal}
\end{equation}
 $D_{\mu\nu}$ is the free gluon propagator, $\sigma$ is the string tension, $S_{\rm min} $ is 
the minimal area enclosed in the {\it distorted} Wilson loop contour (cf. Fig. \ref{figdue}), $C$ is a constant 
and ${\cal P}$ is the perimeter of the loop. The general expression for the minimal surface with contour 
$\Gamma$ is the Nambu-Goto action. In ref.\cite{vel}, it is shown that, at the  order $v^2$ 
of the interaction, the minimal area is equivalent to the area spanned by straight lines connecting 
points at equal time on the quark and antiquark trajectories. For any details we refer to \cite{vel,mod}.
Here, we only present the final result of this minimal area law model (MAL)
\begin{eqnarray}
& &\!\!\!\!\!\!\!\!\!\!\!\!\! V_0(r)  = -{4\alpha_{\rm s} \over 3 r} +  \sigma r + C \,\,\, 
 V_{\rm b}(r) =   {8 \alpha_{\rm s} \over 9 r} - {\sigma \over 9} r\,\,\,
 V_{\rm c}(r)  = - {2 \alpha_{\rm s} \over 3 r} - {\sigma\over 6} r \quad
  V_{\rm d}(r) =  -{\sigma\over 9}  r -{C\over 4}   \nonumber \\
&&\!\!\!\!\!\!\!\!\!\!\!\!  V_{\rm e}(r)  =  -{\sigma \over 6} r \quad
 V^\prime_1(r)  = -\sigma \quad   V^\prime_2(r)  =
  {4\alpha_{\rm s} \over 3 r^2} \quad
 V_3(r) =  4 {\alpha_{\rm s}\over r^3} \quad
 V_4(r)  = {32\over 3}\pi \alpha_{\rm s} \delta^3({\bf r})
\label{rismal}
\end{eqnarray}
and $\Delta V_{\rm a}=0$.
The one gluon exchange term (proportional to $\alpha_{\rm s}$) coincides with the usual Breit-Fermi 
potential. Let us focus on the nonperturbative part (in $\sigma$ and $C$). The $V_1-V_4$ 
spin potentials  (sd) agree with the electric confinement calculation of Eichten and 
Feinberg\cite{eichten} (taking into account the Gromes correction\cite{gromes}).
 The $V_{\rm a}-V_{\rm e}$ velocity potentials (vd)  were first calculated 
in MAL\cite{vel,potlat}. To make clear the physical content of the result, let us write the nonperturbative
contributions in the center of mass system for equal masses. The spin-dependent part is
\begin{equation}
 V_{\rm SD}^{NP}= -{\sigma \over 2 m^2 r} {\bf L}\cdot {\bf S}
\label{sdnp}
\end{equation}
  and corresponds to
 {\it pure Thomas precession},
i.e. the magnetic contribution is zero. The velocity dependent part 
 is 
\begin{equation}
 V_{\rm VD}^{NP}=-{\sigma \over 6} 
{{\bf L^2}\over m^2 r} 
\label{vdnp}
\end{equation}
 and it is proportional to the flux tube angular momentum. Indeed, the same 
vd correction  is obtained in the relativistic flux tube model \cite{olsson} where the energy of the interquark
flux tube is explicitly added to the Hamiltonian\footnote{In the relativistic flux tube model the Lagrangian is composed
by the relativistic energy of the quarks plus the relativistic energy of the flux tube considered as a mechanical 
tube with constant energy density and velocity transversal to the interquark line, see\cite{olsson}.}.  Then, it is clear 
that our gauge-invariant Wilson loop approach has automatically included the energy of the flux tube\footnote{However,
 we are not using any  'mechanical model.'}.\par
The result is even more interesting since {\it it cannot be obtained} within the semirelativistic reduction of a pure 
convolution (i.e. depending only  on the momentum transfer $Q=p_1-p_1^\prime$) Bethe-Salpeter kernel, with any mixture 
of vector and scalar components. In general, a scalar ${1/ Q^4}$ BS kernel is used in order to reproduce 
(\ref{sdnp}), however the velocity dependent correction differ from (\ref{vdnp}), see Sec. 5.  
 On the other hand, the result (\ref{sdnp})-(\ref{vdnp}) is physically transparent. Imagine a quark-antiquark 
pair connected by a chromoelectric flux tube. The magnetic field is zero in the comoving frame and then the spin-interaction
is purely Thomas precession; the electric tube is moving with a transversal velocity and then its energy 
originates the vd term. This is the Buchm\"uller picture. 
We emphasize that the 'scalar-like' character of the spin splitting we obtained is {\it dynamically} generated through 
 {\it the  collective nature of the gluonic degrees of freedom}.
  It is interesting that in \cite{swanson} this same conclusion 
has been obtained working in the framework of the Coulomb gauge Hamiltonian and diagonalizing a sector of the Fock space.
Again, the scalar interaction is effectively generated. However, the Wilson loop approach appears to be simpler and more 
powerful (at least for heavy quarks); the intermediates states that are explicitly considered in the Fock Hamiltonian
approach here are  summed up into the functional formalism. Expressing the potential in terms of 
the Wilson loop we  keep the relevant degrees of freedom. Our special 'trial function' constructed with 
the string interquark operator has selected out the flux tube configuration.\par
The MAL model is in many respects 
  too rough (e.g. the $\langle\langle F_{\mu\nu} F_{\rho\sigma}\rangle \rangle$ v.e.v. turns
 out to be zero and the flux tube turns
out to be infinitely thin) and it may play some role only in the limit of  very large interquark distances
 where an effective relativistic string model description  is expected to  hold.
In the next Sections we discuss more sophisticated assumptions.

\subsection{Gaussian dominance in gluodynamics versus Abelian dominance in infrared QCD}

We need models of the QCD vacuum which  provide us with the nonperturbative behavior of the Wilson loop average.
To this aim we want to exploit all the available lattice information on the mechanism of confinement and 
 the measurements of Wilson loop and field strength v.e. v..  About the mechanism of confinement, 
 at this Conference,    evidence \cite{pol,bali} was presented  that the QCD infrared dynamics
is well approximated by a (dual) Abelian Higgs model. Indeed, in the Abelian  projection the QCD gluodynamics is 
reduced to Abelian fields, Abelian monopoles and charged matter fields degrees of freedom. Lattice 
simulations show Abelian dominance and monopole dominance in the long range features of QCD and condensation
of monopoles in the confined phase. 
For details we refer to \cite{pol,bali}. The measured electric fields and magnetic currents  in the  presence of 
static quark sources are consistent with dual Ginzburg--Landau type of equations. The penetration 
length $\lambda={1/M}$ ($M=$ dual gluon mass) and the correlation length $\xi={1/M_\phi}$ ($M_\phi=$ Higgs mass)
 are measured. It is found $M \simeq M_\phi$ and therefore the QCD vacuum behaves as a dual 
superconductor on the border between type I and type II.  Flux tube solutions (of the type of Abrikosov-Nielsen-Olesen 
vortices) exist and the structure of these solutions is controlled by the penetration and 
the correlation length. However these are lattice measurements and  it is not clear e.g. the relation of 
$\lambda$ and $ \xi$  with the parameters of QCD.\par
On the other side, let us consider the v.e.v. of the Wilson loop. It pays to expand this average 
in terms of fields strength expectation values, by using the non-Abelian Stokes theorem\cite{svm}
\begin{eqnarray}
& & \!\!\!\!\!\!\!\!\!  \langle W(\Gamma)\rangle\equiv \langle \exp
\{ig  \oint_\Gamma dz_\mu A_\mu(z)\} \rangle
=\langle P \exp ig \int_{S(\Gamma)} dS_{\mu\nu}(1) U(0,1)
   F_{\mu\nu}(1) U(1,0)\rangle = 
\label{cum} \\
& & \!\!\!\!\!\!\!\!\!\!\!  \exp \{\sum_{n=0}^\infty {(ig)^n \over n!} 
\int_{S(\Gamma)} dS_{\mu_1\nu_1}(1) \cdots  dS_{\mu_n\nu_n}(n)  
\langle U(0,1)
 F_{\mu_1\nu_1}(1) U(1,0) \cdots
  F_{\mu_n\nu_n}(n)U(n,0) \rangle_{\rm cum}\}
\nonumber
\end{eqnarray}
where $i\equiv x_i$, $S(\Gamma)$ denotes a surface with contour $\Gamma$ and $\langle \dots\rangle_{\rm cum}$ 
stands for the cumulant average\cite{svm}. In principle, the v.e.v. of the Wilson loop is given 
by the sum of all the  cumulants. However, in a recent lattice  investigation
\cite{balinoi}, evidence of the Gaussian dominance in the cumulant expansion of quasi-static Wilson loop 
average was found. Therefore, it follows that 
\begin{equation}
\langle W(\Gamma) \rangle \simeq \exp \{ {-{1\over 2}} \int_{S(\Gamma)} dS_{\mu\nu}(0) \int_{S(\Gamma)} 
dS_{\rho\sigma}(x)
\langle g^2 U(0,x)F_{\mu\nu}(x)U(x,0)F_{\rho\sigma}(0) \rangle \}
\label{gauss}
\end{equation}
is a good approximation. This is  the basic assumption in the (Gaussian) stochastic vacuum model
 \cite{svm} and it was   phenomenologically confirmed by calculation in high energy scattering \cite{dosch}
and quarkonia. Then, the heavy quark interaction is  determined by the two-point field 
strength correlator
\begin{eqnarray}
&& g^2 
\langle U(0,x) F_{\mu\nu}(x) U(x,0) F_{\lambda\rho} (0) \rangle =
  \ {g^2 \langle F^2(0)\rangle \over 24 N_c } 
\bigg\{ (\delta_{\mu\lambda}\delta_{\nu\rho} - 
\delta_{\mu\rho}\delta_{\nu\lambda})(D(x^2) + D_1(x^2)) 
\nonumber \\
&&  + (x_\mu x_\lambda \delta_{\nu\rho} - 
x_\mu x_\rho \delta_{\nu\lambda} 
+ x_\nu x_\rho \delta_{\mu\lambda} - x_\nu x_\lambda \delta_{\mu\rho})
{d\over dx^2}D_1(x^2) \bigg\} .
\label{due}
\end{eqnarray}
In (\ref{due}) the Lorentz decomposition is general and the dynamics is contained in the form factors $D$ and $D_1$.
The function $D$ is responsible for area law and confinement (indeed in QED, due to the  Bianchi identity,  we
have $D=0$). For $D$ and $D_1$
the lattice calculations\cite{balinoi,digiaco}
 gives an exponential long-range decreasing behavior  $\simeq G_2 \exp\{-\vert x\vert /T_g\}$, 
where $G_2\equiv \langle \alpha_s F^2(0)\rangle/\pi $ is the gluon condensate and $T_g\simeq 0.2\,  {\rm fm}$
\footnote{Phenomenological calculation in high energy scattering indicates $T_g\equiv 0.3\div0.35 \, {\rm fm}$\cite{dosch}. }
 is the gluon correlation length. \par
In ref.\cite{abel}, the QCD two--point field strength correlator (\ref{due}) has been related to the dual field 
propagator of the effective Abelian Higgs model describing infrared QCD. In this way the Gaussian dominance
in the Wilson loop average is understood as following from the classical approximation\footnote{Surely valid
in the dual description.} in the dual theory. Moreover, it is possible to relate the QCD parameter $T_g$ and
$G_2$ to the dual parameters.
 In the London limit $T_g$ is identified with the dual gluon mass $M$,
without the London limit the relation is more involved but still $T_g$ is expressed in terms of the dual theory 
parameters. \par 
The conclusion is that we need two parameters $T_g$ and $G_2$ to describe the heavy quark dynamics and indeed 
they are necessary to control the structure of the flux tube. Had we only one parameter, like the string tension 
$\sigma$, we could  encode the information of a constant energy density in the flux tube. However,
the whole structure is important, and also the information about
 the width of the flux tube has to be considered. In the limit of very large interquark distances 
 and in particular dynamical regimes, we can store the relevant information in one parameter, the string 
tension.

\subsection{Vacuum models for the Wilson loop}

 As  shown above, 
 one obtains the $O(v^2)$ quenched quark dynamics in a given vacuum model, 
 simply evaluating the Wilson loop in that model.
 In this way the phenomenological data are put in direct relation with the assumptions  on the QCD vacuum.
 Here we consider three models: Stochastic vacuum model (SVM)\cite{dosch}, 
dual QCD (DQCD)\cite{baker}  and the flux tube model of 
Isgur and  Paton\cite{fluxmod}. We emphasize that these are models of the QCD vacuum, valid at the 
confinement scale.\par\noindent
$\bullet${\it Stochastic vacuum model.}
The stochastic vacuum model\cite{dosch} is based on the idea that the infrared part of the QCD
 functional integral can be approximated by a stochastic process 
with a  converging cluster expansion and a finite correlation length $T_g$. This assumption is well
confirmed by the lattice data. Then, the Wilson loop is given by Eqs.(\ref{gauss})-(\ref{due}) with 
parameters $T_g$ and $G_2$. The behavior of $D$ and $D_1$ and the value of $T_g$ are  taken  from the lattice
measurements, $G_2$ from the phenomenological data. The static potential is
\begin{equation}
V_0(r) \simeq  G_2 \int_0^\infty d\tau
\int_0^r d \lambda (r-\lambda) D(\tau^2 +\lambda^2)
\label{svpot} 
\end{equation}
and the string tension  $\sigma$ emerges as an integral on the $D$ function  $\sigma \simeq G_2 \int_0^\infty d\tau
\int_0^\infty  d\lambda$ $  D(\tau^2 +\lambda^2) $ in the limit ${T_g/ r}\to 0$.
The field distribution between the quark and the antiquark is a flux tube\cite{svmflux}  with 
$r_{MS} \simeq 1.8\, T_g$.  Similarly the sd and vd potential are obtained. I refer to \cite{svm,mod}
for the details. Here we  restrict  the  discussion to the long range limits of the potentials in the three 
vacuum models, (see the final subsection).\par\noindent
$\bullet${\it Dual QCD.}
 DQCD\cite{baker}  is a concrete 
 realization of the Mandelstam t'Hooft dual superconductor mechanism of confinement.
It describes the QCD vacuum as a dual superconductor on the border between type I and II. 
The Wilson loop approach supplies with  a simple method to connect 
averaged local quantities in QCD and in dual QCD (=dual gluodynamics).
For large loops we assume\cite{baker,dual} 
\begin{equation}
\langle W(\Gamma) \rangle \simeq \langle W(\Gamma)\rangle_{Dual}= 
{\displaystyle{\int \!{\cal D} {\cal C} \,  {\cal D} {\cal B} \,
\exp \left[ i \int dx  L(G_{\mu\nu}^{\rm S}) \right] }
\over \displaystyle{
\int \!{\cal D} {\cal C}\, {\cal D} {\cal B} \,  
\exp \left[ i \int dx  L(0) \right]  }}
\label{deq}
\end{equation}
where $C_\mu$ are the dual potentials, $G_{\mu\nu}$ the dual field strengths, ${\cal B}_i$ 
the monopole fields,
 ${ L} $
the effective dual Lagrangian in the  presence of quarks
$
{ L}(G_{\mu\nu}^{\rm S}) = 
2~{\rm Tr} 
\{ - {1\over 4} {\cal G}^{\mu\nu} {\cal G}_{\mu\nu}$ 
$+ {1\over 2} ({\cal D}_\mu {\cal B}_i)^2  \} 
- U({\cal B}_i)
$, and 
$U({\cal B}_i)$  the Higgs potential. 
The quark sources moving around the Wilson loop are  inserted via the Dirac string tensor  
$
G_{\mu\nu}^{\rm S}(x)  = g \,\epsilon_{\mu\nu\alpha\beta} 
\int ds \int d t \, y^{\prime \alpha} \dot{y}^\beta$ 
$  \delta(x-y)$.
From (\ref{deq}) it follows
\begin{equation}
g \langle\!\langle F_{\mu \nu} (z_j)\rangle\!\rangle
= {2\over 3} g \, \varepsilon_{\mu\nu \rho\sigma}\langle\!\langle 
G^{\rho\sigma} (z_j) \rangle\!\rangle_{Dual}
\label{valdual}
\end{equation}
and  it is possible to relate the averaged values of local quantities in QCD and in the dual theory. Then
 Eqs.(\ref{vsd})-(\ref{vd})  produce the $O(v^2)$ interaction in the dual formalism, for details
 cf.\cite{dual}. We discuss only the long range limits in the final subsection. 
The nonperturbative parameters are  the  v.e.v. of
 the Higgs field, ${\cal B}_0$, and $\kappa$, the  
 coupling constant of the Higgs potential  (from these the penetration length and the correlation length can
 be constructed).  The flux tube configurations  ($\simeq$ Abrikosov-Nielsen-Olesen vortex) with finite 
 $r_{MS} \sim {1/ M}$, $M=$ mass of the dual gluon, arise\cite{fluxdual}
 from the numerical solution of the classical
dual Ginzburg-Landau type of equations obtained from ${ L}$.
It is the presence of the Higgs field which confines transversally the energy distribution in a flux tube.
 Again 
$\sigma$ comes from the integration on a function  exponentially decreasing
approximately   with $M$. \par\noindent
$\bullet${\it Isgur-Paton flux tube model.}
This model is extracted from the strong coupling limit of the QCD lattice Hamiltonian.
 A N-body discrete string-like model Hamiltonian describes the gluonic degrees of freedom.
The limit $N\to \infty$ corresponds to a localized string with an infinite number of degrees of freedom;
the radius of the flux tube is proportional to ${1/N}$.
  For any detail
we refer to the talk of Paton in these Proceedings.  In a recent paper\cite{swanson}, the sd potential
has been  obtained in this model.

\subsection{ Potentials and flux tube structure}

It is interesting to consider the potentials obtained using these  different 
vacuum models in the limit of large interquark distance $r$.\par
\leftline{$\bullet$\it SVM}
\begin{equation}
V_0(r) =   \sigma r  + {1\over 2}c^{(1)} \sigma\, T_g
 - c\, \sigma\, T_g  , \quad
{d\over dr}V_1(r) =  -\sigma  + {c\, \sigma\, T_g\over r} \>, \quad
{d\over dr}V_2(r)  =  {c\, \sigma\, T_g \over r}  \>,
\label{limsvm}
\end{equation}
$V_3$ and $V_4$ fall off exponentially as $\exp\{-{r/T_g}\}$ and 
\begin{eqnarray}
& &\!\!\!\!\!\!\!\!\!\!\!\!\!\!  V_{\rm b}(r) =  -{\sigma\over 9}  r
  -{2\over 3}{d\, \sigma\, T_g^2 \over r} 
+ {8\over 3}{e\, \sigma\, T_g^3\over r^2}\>,
\quad\quad
V_{\rm c}(r)  =  -{\sigma\over 6}  r
 -{d\, \sigma\, T_g^2\over r} + {2\over 3}
{e\, \sigma\, T_g^3\over r^2}\>,
\label{vdsvm}\\
& & \!\!\!\!\!\!\!\!\!\!\!\!\!\!\!  V_{\rm d}(r) =  -{\sigma\over 9} r  + {c\over 4}\,\sigma\, T_g
 - {c^{(1)}\over 8}\sigma\, T_g 
+ {d\over 3}{\sigma\, T_g^2\over r} - {2\over 9}{e\, \sigma T_g^3
 \over r^2}\,\,\,
V_{\rm e}(r)   =
  -{\sigma\over 6}  r  + {d\over 2}{\sigma\, T_g^2\over r} 
- {e\over 3}{\sigma\, T_g^3\over r^2}.
\nonumber
\end{eqnarray}
\leftline{$\bullet$\it DQCD}
\begin{equation}                                       
 V_0 (r)  =  \sigma r - (C+C^{(1)}) {\sigma\over M}\quad\quad  {d \over dr} V_1(r) =  -\sigma 
+   C {\sigma \over M} {1 \over r}\quad\quad  
{d \over dr} V_2(r) = C {\sigma\over M}{1 \over r}
\label{limdq}
\end{equation}
and
$V_3$ and $V_4$ fall off exponentially as $\exp\{-M r\}$.\par
\leftline{$\bullet$\it I-P flux tube model}
\begin{equation}
{d\over dr} V_1(r)= -\sigma 
\quad {d \over dr}V_2(r)=  \lim_{N\to \infty} {\sigma\over N}
\label{limflux}
\end{equation}
and 
$V_3$ and $V_4$ are $\simeq {1/N}$.
The $c,c^{(1)},d,e$, $C, C^{(1)}$ 
above are numerical constants known in terms of the 
parameters of the models.
We learn that
\begin{description}
\item{1)} The  gluon  correlation length $T_g$ in SVM
 has the same role as
the inverse of the dual gluon mass $M$ in DQCD and $T_g/r$ as the inverse of the
number of degrees of freedom $ N $ in the I-P flux tube model. All these quantities act like a
correlation length. From now on, we use $T_g$ to refer to any of them. 
\item{2)} The MAL results are  completely contained in all these
 models\footnote{For the vd potential in DQCD see \cite{dual}.}.
The MAL  describes the limit of large interquark distances or better the limit
${T_g/r}\to 0$ which is also the limit in which a potential exists, cf. Sec. 6, or the string limit.
Indeed, all the corrections to the MAL  are 
proportional to the  gluon correlation length or equivalently to the
 width  of the flux tube: the form of these corrections is {\it the same}  in all
these flux tube-like  models (models that predict the flux tube 
structure).
\item{3)} The magnetic  interaction is zero only in the limit $T_g/r \to 0$.
In the intermediate distance  region is different from zero as one  can see from the value of $V_2$.
Then a purely scalar effective interaction emerges only in the long range limit.
\item{4)}  $T_g$ controls the width and the shape 
 of the flux tube as well as the validity of the potential description.
Physically it should correspond to the size of the color domain as well as
the fluctuation of the color fields.
\end{description}
We conclude that {\it the results for the heavy quark interaction are essentially the same for any flux-tube 
model of the QCD vacuum}.

\section{Lattice calculations, flux tube models and ${1/Q^4}$ models}
\begin{figure}[htb]
\makebox[0.5truecm]{\phantom b}
\put(5,-50){\epsfxsize=6.7cm \epsfbox{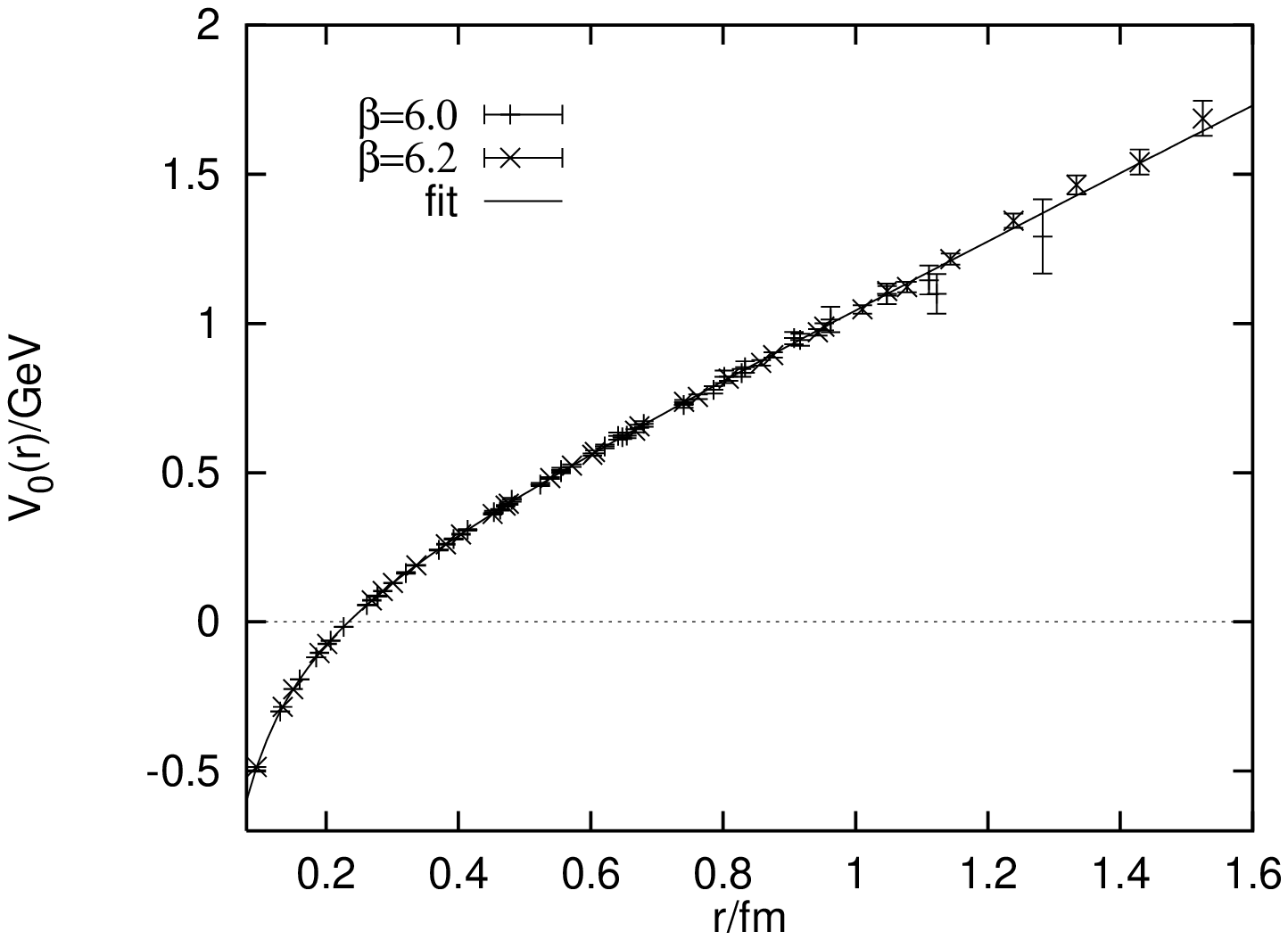}}
\put(220,-50){\epsfxsize=6.7cm \epsfbox{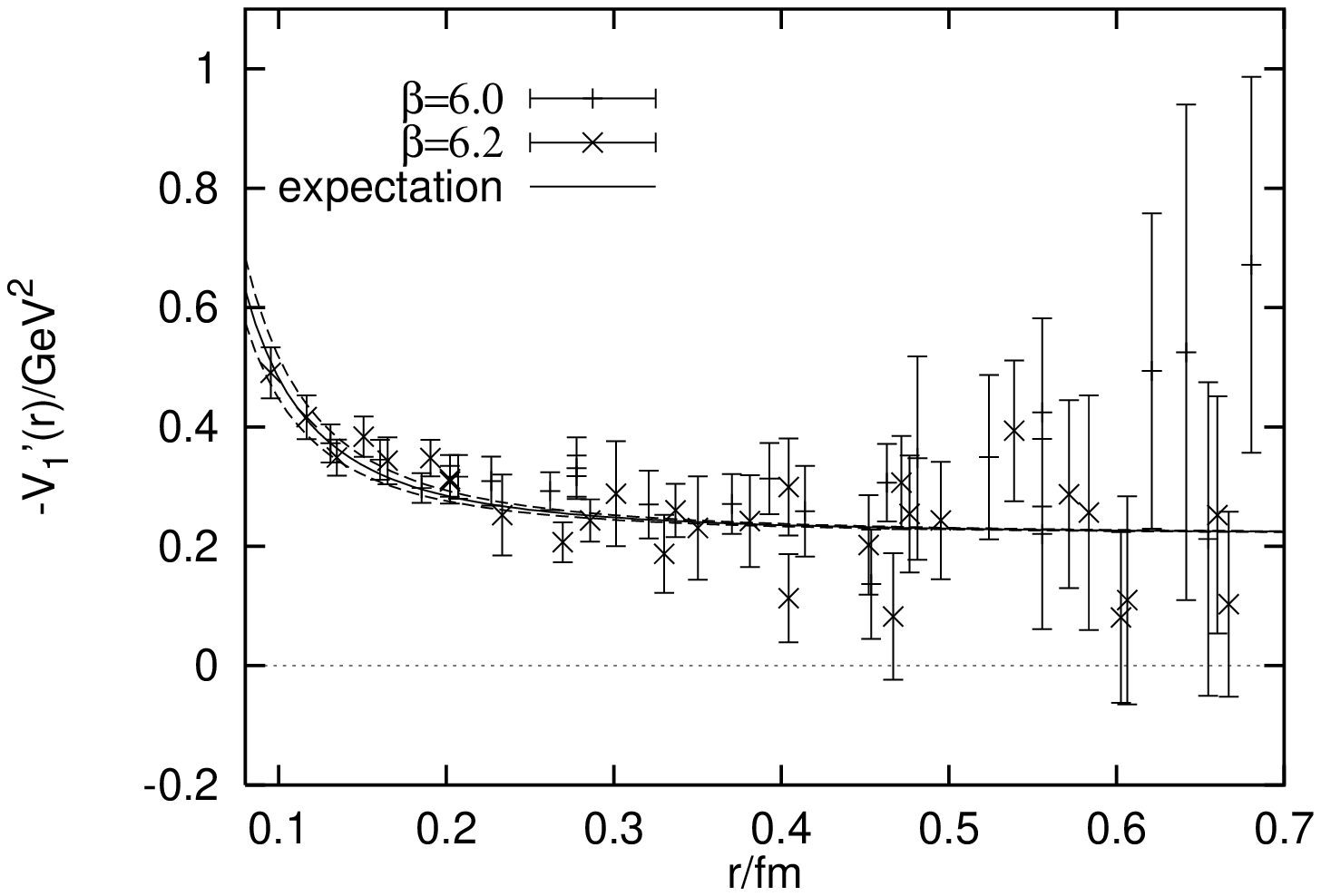}}
\caption[x]{$V_0$ and $V_1^\prime$ potentials from \cite{balipot}.}
\label{fig:01}
\end{figure}
\vspace{0.3cm}

\begin{figure}[htb]
\makebox[0.5truecm]{\phantom b}
\put(5,-50){\epsfxsize=6.7cm \epsfbox{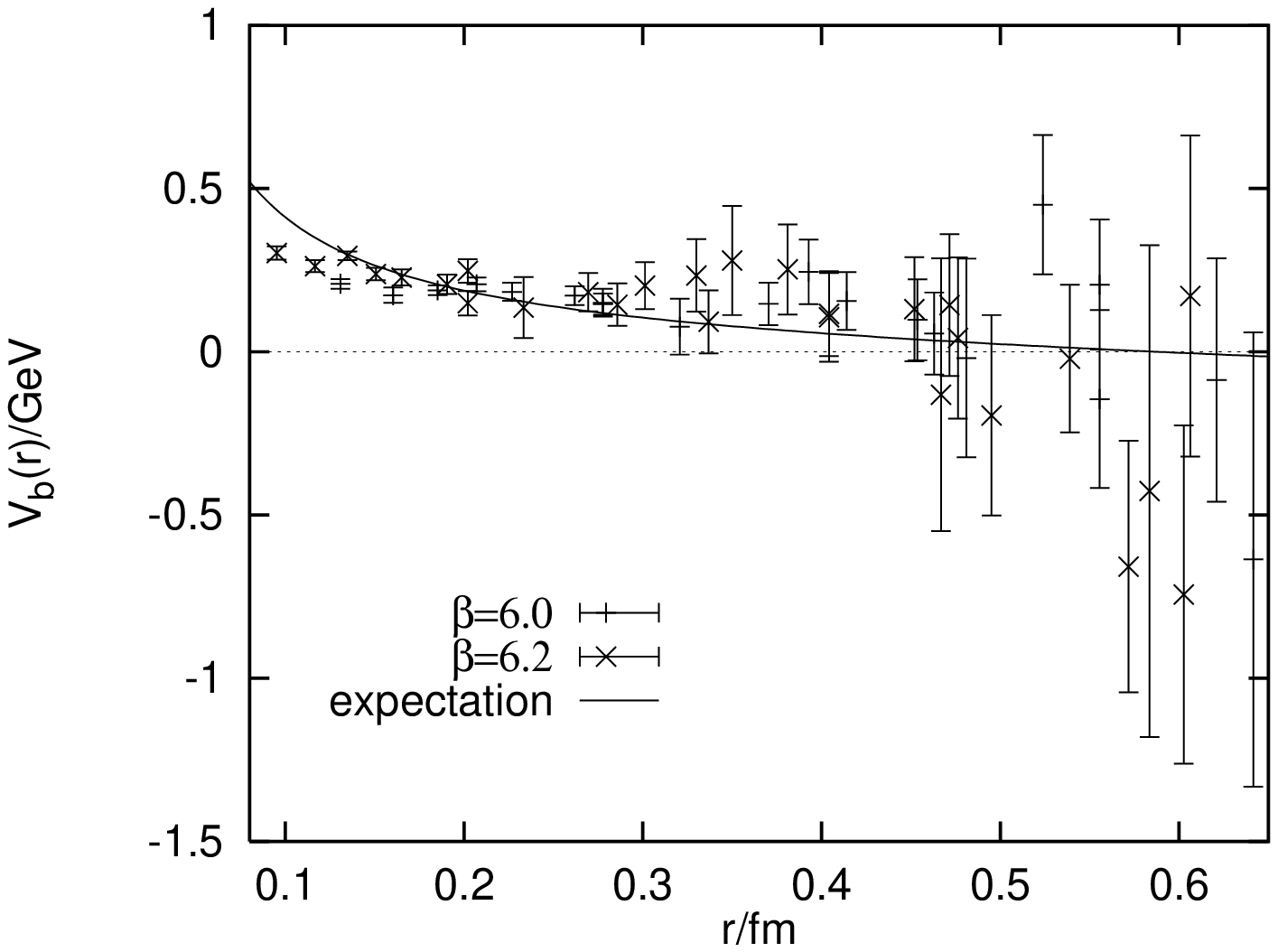}}
\put(220,-50){\epsfxsize=6.7cm \epsfbox{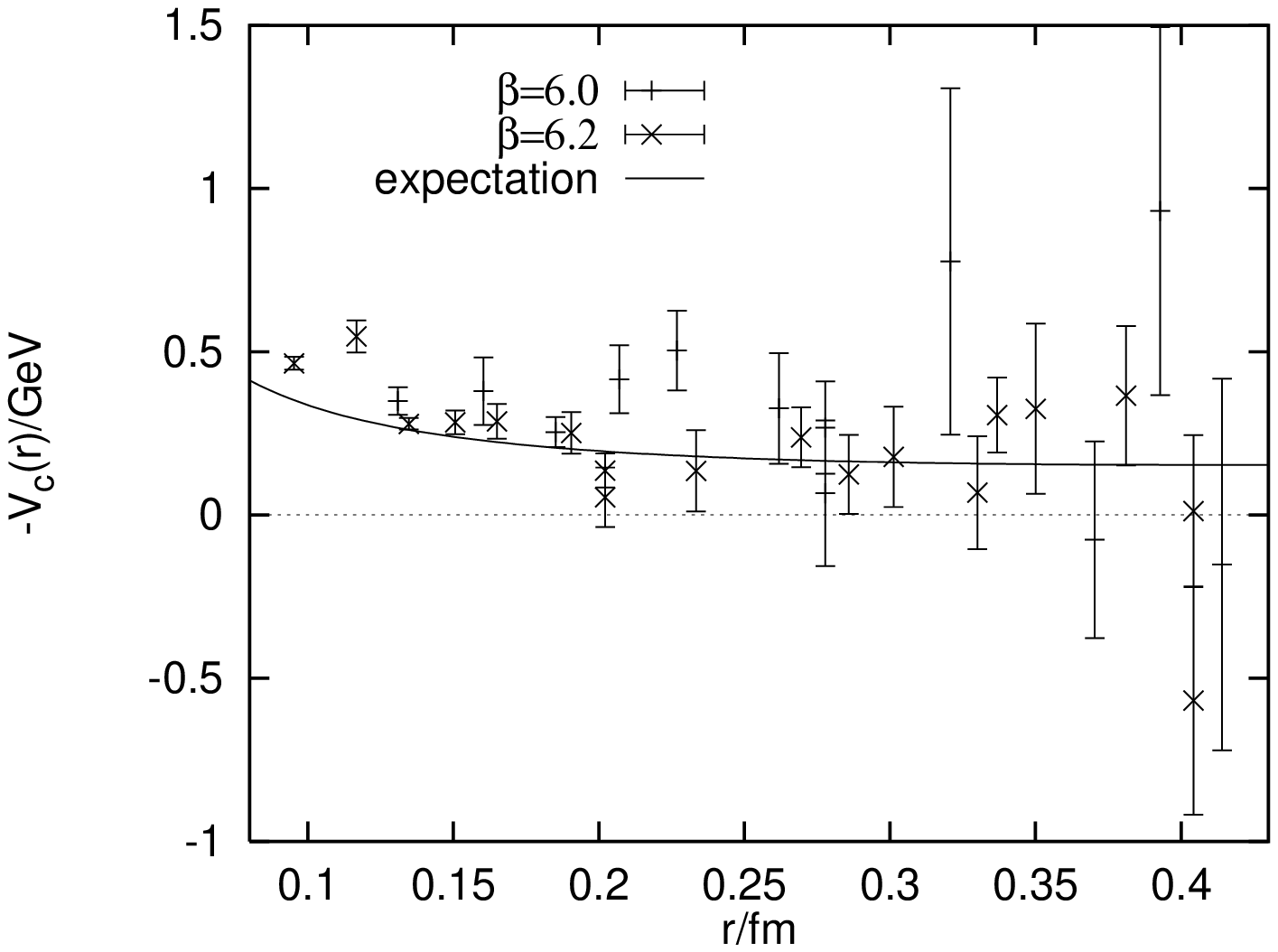}}
\caption[x]{$V_{\rm b}$ and $V_{\rm c}$ potentials from \cite{balipot}.}
\label{fig:12}
\end{figure}
\vspace{0.3cm}

\begin{figure}[htb]
\makebox[0.5truecm]{\phantom b}
\put(5,-50){\epsfxsize=6.7cm \epsfbox{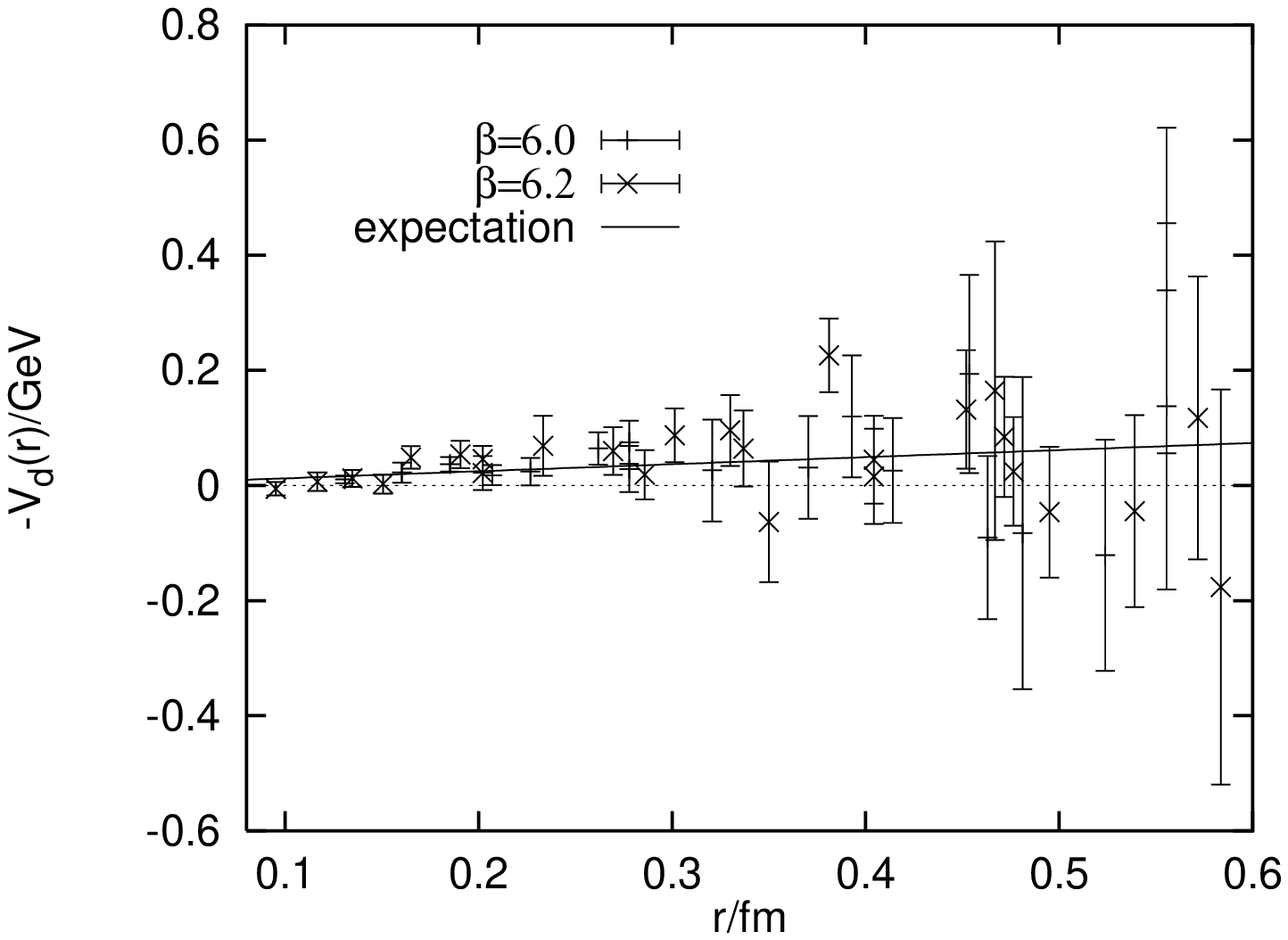}}
\put(220,-50){\epsfxsize=6.7cm \epsfbox{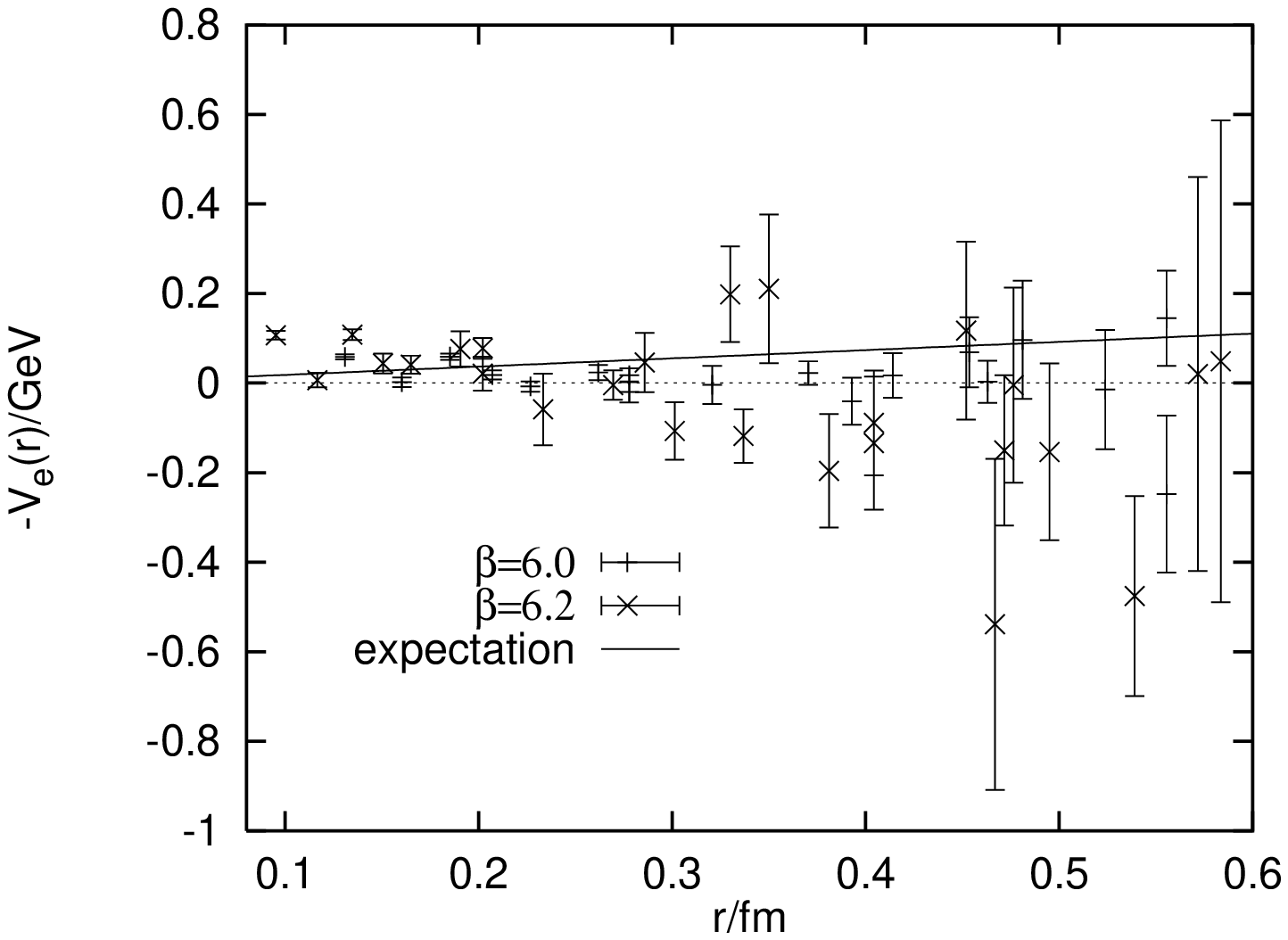}}
\caption[x]{$V_{\rm d}$ and $V_{\rm e}$ potentials from \cite{balipot}.}
\label{fig:34}
\end{figure}
\vspace{0.3cm}
An entire class of confining models for the heavy quark dynamics
 comes from the semirelativistic reduction 
of a  pure convolution Bethe-Salpeter  kernel (see\cite{report} and refs. therein)
\begin{equation}
I(Q^2)=  \gamma_1^\mu  \gamma_2^\nu P_{\mu\nu}  J_v(Q) + J_s(Q)
\label{kernel}
\end{equation}
where $J_v$ contains the one gluon exchange, $P$ is a 
 factor depending on the gauge and $J_s$
has   an infrared  behavior like  ${1/ {\bf Q}^4}$
(or ${1/Q^4}$, $Q$ being the momentum transfer) chosen in order to reproduce the linear behavior of the static potential.
 To reproduce $V_1^\prime$ then      $J_s$  has to be  scalar.
These models produce:
\begin{eqnarray}
& &\!\!\!\!\!\!\!\!\!\!\!\!\!\!  V_0    = -{4\alpha_{\rm s} \over 3 r} +  \sigma r + C \quad 
 V_{\rm b}(r) =   {8\alpha_{\rm s} \over 9 r} +0/{2\sigma \over 3} \quad
 V_{\rm c}(r)   =  - {2\alpha_{\rm s} \over 3 r} - {\sigma \over 2} r/0
\quad 
 V_{\rm d}(r) =  -{\sigma \over 2}  r -{C\over 4}   \nonumber \\
& & \!\!\!\!\!\!\!\!\!\!\!\!\!\!  V_{\rm e}(r)  =  0 \quad 
 V^\prime_1(r)   =  -\sigma \quad   V^\prime_2(r)  =
  {4\alpha_{\rm s} \over 3 r^2} \quad
 V_3(r) =   {4 \alpha_{\rm s}\over r^3} \quad
 V_4(r)  = {32\over 3}\pi \alpha_{\rm s} \delta^3({\bf r})
\label{quares}
\end{eqnarray}
The two results in Eq. (\ref{quares}) correspond respectively
 to the instantaneous form of the kernel ${1/{\bf Q}^4}$ or to 
the inclusion of  retardation corrections as in \cite{faustov}.
Notice that: 1) These models are constructed in order to reproduce $V_1$ as given in Eq. (\ref{rismal}).
2) They cannot reproduce the behavior of the $V_{\rm VD}^{NP}$ (\ref{vdnp}) 
  (even  adding retardation 
corrections or allowing  a mixture of vector and scalar confining kernels\cite{faustov}).
3) The ${1/ Q^4}$ 
propagator {\it does not avoid the spreading of the flux tube} between the quarks
{\it even if the  integrated energy increases linearly with the distance} between the
quarks.\par
 We conclude that a linear confining behavior
with the correct spin-orbit   {\it is not sufficient to characterize} 
 the heavy quark dynamics at the order $O(v^2)$. On the other hand the $v^2$ corrections to the potential are connected
to v.e.v. of field strengths insertion in the Wilson loop and therefore 
 to  the energy density and to the interquark flux tube structure. {\it The $1/Q^4$ models fail to take into account the dynamical 
nature of the glue}.
In Figs. 3-5  we present the latest lattice results for the potentials\cite{balipot}. The  lines 
are the MAL predictions (\ref{rismal}).
The static potential is clearly linear for $r > 0.2\, {\rm fm}$ (recent results indicate that 
the string breaking happens for $r> 0.2\, {\rm fm}$ \cite{bali}). The nonperturbative behavior of $V_1^\prime$ is
 consistent with the result (\ref{rismal}) and (\ref{quares}). Unfortunately, the lattice data for the vd 
potential are not yet sufficiently accurate to discriminate between the flux tube model and the 
${1/ Q^4}$ predictions. Notice that an accurate measurement of $V_{\rm e}$ provides a good test since it is predicted to be zero in ${1/ Q^4}$ 
models. \par
Phenomenological applications to the bottomonium and charmonium spectrum show that the vd corrections
of the type (18)  do not induce  the deviations from the data originated by the  vd corrections (32)\cite{bp}.

\section{On the validity of the potential description}
In the previous sections  we assumed that a potential description holds. 
This is not always the case. We know from the Lamb-shift in QED that at some order 
of the perturbative expansion the ultrasoft photons start  giving  a contribution
to the energy not of a potential type. This is certainly true also in QCD but here 
we cannot rely only on the perturbative calculation. Then, the situation becomes 
more interesting and again the flux tube configuration plays a role. We can extract a 
potential as far as the adiabatic approximation holds, i.e., as far as the gluon time scale 
is short  compared to the orbital period of the quarks $T_g\ll T_q$. 
In this case we are describing a $ q\bar{q}$ pair moving in the adiabatic potential 
corresponding to the ground state energy of the gluon field, i.e. the ground state
flux tube. The next adiabatic surface describes an excited state of the glue:
{\it  an hybrid}
\cite{michael}. If these adiabatic surfaces are well separated,  still each of them can 
be described with a potential.  The situation is controlled by the gluon correlation
 length $T_g$ which is proportional  to the this difference. The quark dynamics is simply
given  in terms of the two-point correlator (\ref{due}) whose nonperturbative exponential
decreasing behavior is controlled by $T_g$. In the case in which $T_g$ is small with 
respect to the other physical scales of the problem we recover the potential description,
in the opposite limit we recover the sum rules description with local condensates
 \`a la Leutwyler-Voloshin
\cite{marquand}. Therefore there is no discrepancy between these two descriptions.
However, the $c\bar{c}$ and $b\bar{b}$ systems lie in the situation in which 
$T_g$ is small compared to the other scales. Just for the ground state of the
 $ b\bar{b}$  both descriptions can be undertaken.\par
Recently a new effective theory  has been obtained from NRQCD integrating out  the relative momentum scale.
It is called potential non relativistic QCD (pNRQCD) and was presented at this Conference by J. Soto
\cite{pnrqcd}. The theory describes ultrasoft gluons 
and  the matching coefficients are the potentials. pNRQCD  could be a powerful tool 
to understand the relation between ultrasoft degrees
 of freedom and nonpotential nonperturbative glue.

\section{Confinement, flux tube and chiral symmetry}

As soon as the light quarks enter the game we are left without any  expansion parameter to 
develop any well--founded calculation of the quark dynamics. Even in the simplest case,
a meson formed by a light and a heavy quark, we are far away from the intuitive description 
developed for the heavy--heavy bound states.
Indeed, in this case the heavy quark is surrounded by    the  fuzzy cloud  of the light degrees of freedom.
This picture is naively completely different from a flux tube picture. Moreover  chiral symmetry breaking 
should appear.
Here, we address the issue  whether the Wilson loop formalism still helps to understand the quark dynamics.
If this is the case we are in the position to  study  the  interplay  between
confinement and chiral symmetry breaking. \par
Let us consider the simplest system, the heavy-light bound state.
Heavy-light systems are understood in terms of heavy quark effective theory (HQET), but HQET
 cannot  predict the spectrum of the  'brown muck'
since this is determined by the dynamics of strong QCD.
 In the limit  $m_Q \gg \Lambda_{QCD}$, HQET organizes a systematic expansion of the physical quantities in terms 
of $\Lambda_{\rm QCD}/ m_Q$ and $\alpha_{\rm s}(m_Q)$. The meson mass is given as 
\begin{equation}
m_{M}= m_{Q}+ \bar{\Lambda} 
+ O \left( {1\over m_Q} \right) \, \,{\rm corrections}.
\label{hqet}
\end{equation}
The parameter $\bar{\Lambda}$ can be fixed on the data  but its
 actual  calculation needs a dynamical input.
 In the no-recoil limit $m_Q\to \infty$,
 the usual choice is to take a   scalar  Dirac equation to get the $\bar{\Lambda}$ parameter as an
eigenvalue. The motivation is  reproducing the spin-orbit splitting and the fact that this kind of 
equation is mathematically well behaving.
There are a number  of reasons against this choice:
\begin{itemize} 
\item{} It breaks explicitly chiral symmetry.
\item{} We have shown that scalar confinement arises effectively for heavy quark 
interaction due to  the  collective nature of the nonperturbative gluonic degrees of freedom.
\end{itemize}
In the next section we address the problem in the  Wilson loop formalism.

\subsection{Gauge-invariant approach to heavy-light quark systems}

\begin{figure}[htb]
\vskip -0.1truecm
\makebox[1truecm]{\phantom b}
\put(0,-50){\epsfxsize=8.0cm \epsfbox{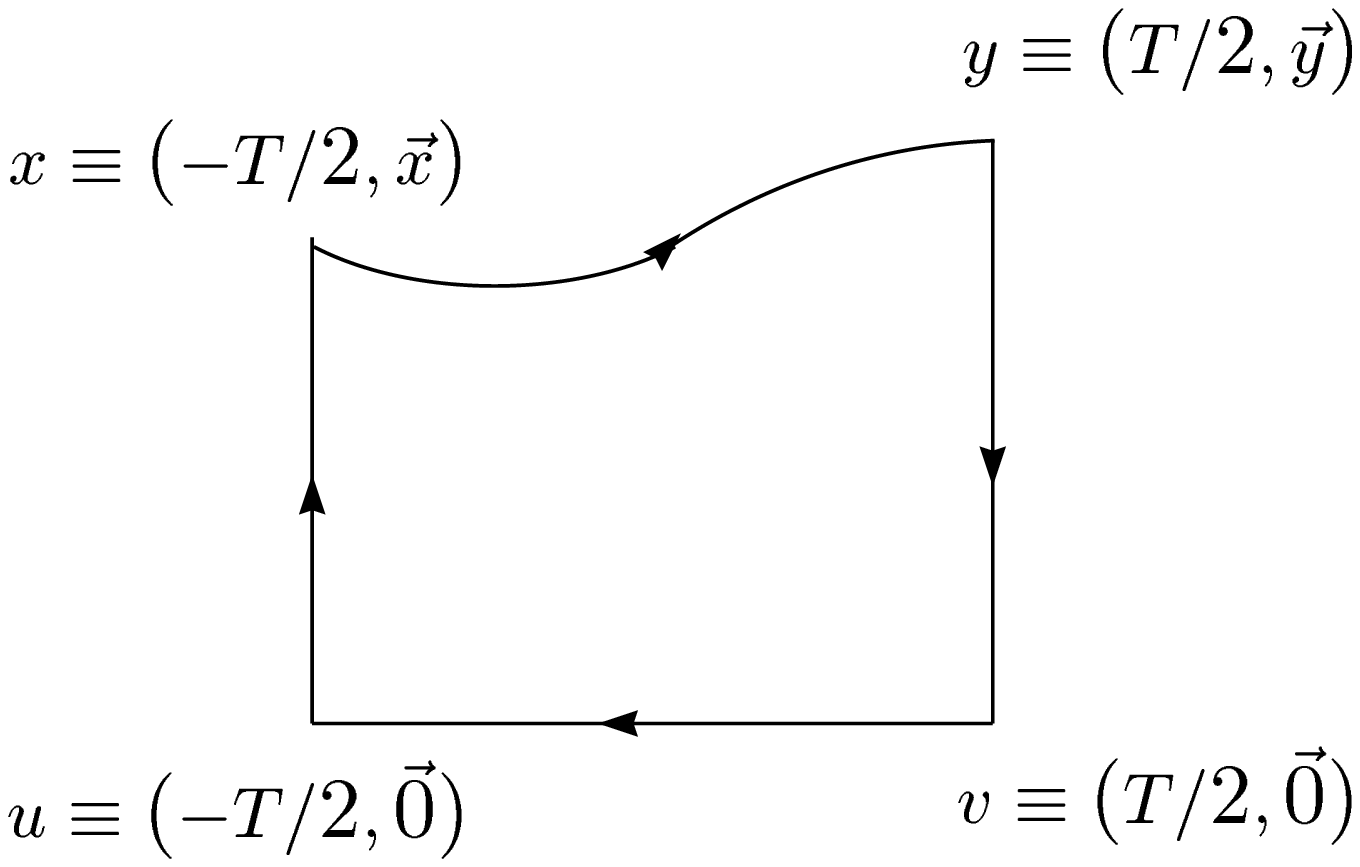}}
\put(220,70){\epsfxsize=6.0cm \epsfbox{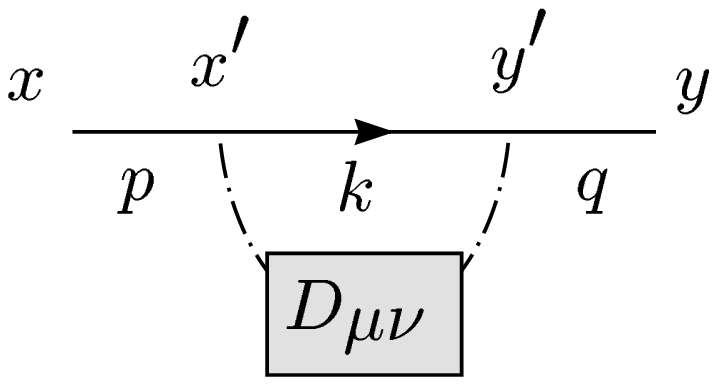}}
\vskip -3truecm
\caption[x]{The Wilson loop in the static limit of the heavy quark and the interaction kernel $K$.}
\label{figundici}
\end{figure}

We start from the gauge-inva\-riant quark-antiquark
 Green function in the Fey\-nman-Schwi\-nger re\-pre\-sen\-tation
\cite{dirac,bs}:
\begin{eqnarray}
&~&\!\!\!\!\!\!\!\!\!\!\!\!\! G(x,u,y,v) =
{1\over 4} \Bigg\langle {\rm Tr}\,{\rm P}\, 
(i\,{D\!\!\!\!/}_{y}^{\,(1)}+m_1)\, 
\int_{0}^\infty dT_1\int_{x}^{y}{\cal D}z_1
e^{\displaystyle - i\,\int_{0}^{T_1}dt_1 {m^2+\dot z_1^2 \over 2}   }
\nonumber\\
&~&\times
\int_{0}^\infty dT_2\int_{v}^{u}{\cal D}z_2
e^{\displaystyle - i\,\int_{0}^{T_2}dt_2 {m^2+\dot z_2^2 \over 2}   }
e^{\displaystyle ig \oint_\Gamma dz^\mu A_\mu(z)}
\label{feyschwi}\\
&~&\times 
e^{\displaystyle i\,\int_{0}^{T_1}dt_1 {g\over 4}\sigma_{\mu\nu}^{(1)}
F^{\mu\nu}(z_1)}
e^{\displaystyle i\,\int_{0}^{T_2}dt_2 {g\over 4}\sigma_{\mu\nu}^{(2)}
F^{\mu\nu}(z_2)} 
(-i\,\buildrel{\leftarrow}\over{D\!\!\!\!/}_{v}^{\,(2)} + m_2) \Bigg\rangle . 
\nonumber
\end{eqnarray} 
Again the dynamics is contained in the Wilson loop, that now looks like
 Fig. \ref{figundici}.
 We can 
exploit the symmetry of the situation, taking the  modified coordinate gauge 
$ A_\mu(x_0,{\bf 0}) = 0, \,  x^jA_j(x_0,{\bf x}) = 0 $  ($A_0(x)=\int_0^1 d\alpha x^k F_{k0}(x_0,\alpha {\bf x})$,
 $ A_j(x)= \int_0^1 d\alpha \alpha x^k F_{kj}(x_0,\alpha {\bf x}))$ in which 
\begin{equation}
W(\Gamma) =    
{ Tr \,} { P\,} \exp \left\{ ig \int_x^y dz^\mu A_\mu (z) \right\}.
\label{wilh}
\end{equation}
At this point, at variance from the heavy quark case, we have to make a  model dependent assumption: 
we consider  still valid the  dominance of the bilocal correlator with a finite correlation length.
Under this assumption we obtain a Dirac like equation for the light quark  
in presence of the heavy quark.
Approximating  the Wilson loop as 
\begin{eqnarray}
\langle W(\Gamma) \rangle &=& 
\exp \left\{ - {g^2\over 2} \int_x^y dx^{\prime\mu} \int_x^y dy^{\prime\nu}
D_{\mu\nu}(x^\prime,y^\prime) \right\},\label{proph}\\
D_{\mu\nu}(x,y) &\equiv& 
x^ky^l\int_0^1 d\alpha \, \alpha^{n(\mu)} \int_0^1 d\beta\, \beta^{n(\nu)}
\langle
 F_{k\mu}(x^0,\alpha{\bf x})F_{l\nu}(y^0,
-\beta{\bf y})\rangle \nonumber
\end{eqnarray}
where $n(0) = 0$ and $n(i) = 1$, we obtain the Dirac--like equation for the light quark propagator
\begin{equation}
S_D = S_0 + S_0 K S_D
\label{diracl}
\end{equation}
with a kernel
$K(y^\prime,x^\prime) \equiv \gamma^\nu 
S_0(y^\prime,x^\prime) \gamma^\mu D_{\mu\nu}(x^\prime,y^\prime)$,
 given diagrammatically in Fig. \ref{figundici}.
 Notice that\cite{dirac}: 1) $K$ {\it is not a translational invariant quantity}: the coordinate gauge breaks 
explicitly this symmetry in the propagator. Physically this is due to the presence of the heavy quark.
Indeed Eq. (\ref{diracl}) is an integral equation for the light quark propagator in the field of the heavy quark.
2) The kernel depends on $D_{\mu\nu}$ which in turns is given in terms of the two-point correlator (\ref{due}).
Then, the heavy-light dynamics is controlled by the same two parameters controlling the heavy-heavy dynamics, $T_g$ and $G_2$.
3) The problem has many relevant scales:  the light  mass $m$,
 the correlation length $T_g\sim \Lambda_{QCD}$,
 the characteristic energy and momentum of the bound states. We have different dynamical regimes in dependences on the relative 
values of these scales. Let us study the various situations. 
In the following we consider only the nonperturbative dynamics\cite{dirac}.
\begin{itemize}
\item{} {\bf Potential Case:  $m>{1/ T_g} > p_0-m, {\bf p}, {\bf p}-{\bf q}$.} 
We neglect the negative energy states  and expand the kernel $K$ 
 in $m$. We  obtain: 
\begin{equation}
\!\!\!\!\!\!\!\!\!\!\!\! V(r)\sim G_2 \big \{ \int_{-\infty}^{+\infty} d\tau \int_0^r d \lambda(r-\lambda) 
D(\tau^2+\lambda^2) + {{\bf \sigma}\cdot {\bf L} \over 4 m^2 r} 
\int_{-\infty}^{+\infty} d\tau \int_0^r d \lambda 
\left( {2 \lambda\over r} - 1\right) D(\tau^2+\lambda^2)\big\}
\label{poth}
\end{equation}
which coincides in the limit of large $r$ with the Eichten-Feinberg 
potential (\ref{rismal}) 
with   $\sigma$ as defined in the SVM section.
We emphasize that the Lo\-rentz structure  which gives origin to the negative 
sign in front of the spin-orbit potential (hence to the Thomas precession term)  is in our 
case {\it not simply a scalar}  ($K\simeq \sigma \, r$). 
\item{}{\bf Sum Rules case: $({1/T_g}<p_0-m$, ${1/T_g}<m)$.}
We get  the well-known  Shifman, Vainshtein
 and Zakharov result for the heavy quark condensate 
$$
\langle \bar{Q} Q \rangle  = 
-\int {d^4p \over (2\pi)^4}\int {d^4q \over (2\pi)^4} 
{\rm Tr} \left\{ S_0(q)K(q,p)S_0(p) \right\} = -{1\over 12} 
{\langle \alpha F^2(0) \rangle \over \pi m}.
$$ 
\item{\bf $D_s$ and $B_s$ case: ${1/T_g} >m$.} 
 For $D_s$ and $B_s$ one can still assume that the propagator inside 
the kernel is free and solve the equation to get the spectrum.
\item{\bf $D$ and $B$ case: $  m\ll {1/T_g}$.} 
The nonlinear
equation\cite{simonov,cond} 
\begin{equation}
S_D=S_0+S_0 K(S_D)S_D
\label{schwind}
\end{equation}
has to be solved with Schwinger--Dyson like techniques. 
Notice that in the limit  $m\to 0$ the form of the 
kernel $K$ is such that the interaction turns out to be {\it chiral symmetric}.
In Ref.\cite{cond} a solution of
 the gap equation corresponding to Eq. (\ref{schwind}) 
(with a simplified form of the interaction) was found. In that case it was possible
 to disentangle the translational invariant part 
of the interaction (the self-energy) from the non-invariant part (the bound state interaction). In this way 
one  obtains  the light quark condensate
\begin{equation}
\langle 0\vert \bar{q} q \vert 0\rangle \sim -   T_g G_2 
\label{cond}
\end{equation}
and the heavy-light bound state energy spectrum. The result (\ref{cond}) looks appealing: it establishes a connection
between the gluon condensate and the light quark condensate. The connection is possible since the non-local 
gluon condensate has introduced into the game a finite correlation length $T_g$. The same quantity that 
controls the width of 
the flux tube as well as the validity of the potential description in the case of heavy quarks.
\end{itemize}

 In conclusion  we have reached 
some  insight in the interplay between confinement and chiral 
symmetry breaking, obtaining for the first time an unified framework
 for  sum rules, potential models and  chiral symmetry breaking  studies.

\section{Gauge-invariant approach to light-light quark systems}

The starting point is again Eq. (\ref{feyschwi}). The dynamics is contained in the distorted Wilson loop of Fig. 2.
 In this situation it does not exist a  choice of the gauge like  the  modified coordinate gauge in the 
heavy-light case, i.e. it is not possible to get rid of both the final Schwinger strings.
A result for the nonperturbative BS kernel can still be obtained in the form of effective diagrams\cite{bs}
but gauge-invariance is lost. Even recovering the potential from such a kernel appears to be problematic.

\section{Conclusions}
We have shown  that the 
 $O(v^2)$ heavy quark interaction is controlled only by the Wilson loop 
behavior: heavy quarks are a nice laboratory to test QCD vacuum 
models with respect to lattice and phenomenological data. On the other hand,
 the Wilson loop  is a good approach to the quark dynamics: it keeps 
 automatically the relevant degrees of freedom and includes naturally the 
 flux tube configurations.  
The description of the heavy quark interaction needs two nonperturbative parameters 
$T_g$ and $G_2$: these give origin to $\sigma$ only in particular
dynamic situations. $T_g$ controls the  width and the shape of the flux tube, as well
as the validity of the potential description.  Any model of the QCD vacuum 
reproducing the interquark flux tube structure gives the same prediction for the nonperturbative 
heavy quark interaction. On the other hand, the class of ${1/Q^4}$ models give definite 
different predictions.  \par
 Moreover, the Wilson loop formalism appears to be useful in selecting the relevant 
configurations also in situations where  an intuitive flux tube picture does not 
exist: e.g. the heavy-light system. In this way  the heavy-heavy and heavy-light systems  are understood 
in terms of the same  parameters ($T_g,G_2$). These parameters are measured on the lattice and are 
also connected to the parameters of the dual superconductor mechanism.\par
Among the open problems/work in progress   we do  list:
extension of  the analytic calculations to the higher order corrections
 in NRQCD  for the heavy quark interaction; inclusion of the  non-potential contribution
(pNRQCD?); treatment in this formalism of systems involving more than one light quark; unquenching.

\section{Acknowledgements}
The author acknowledges the support of the European Community, Marie Curie fellowship, TMR Contract
 n. ERBFMBICT961714.     

\end{document}